\documentclass[PRL,aps,twocolumn,superscriptaddress,10pt,showpacs]{revtex4-1}
\usepackage{graphicx}
\usepackage{epstopdf}
\usepackage{color}
\usepackage{amsmath,amssymb}
\usepackage{fixmath}

\usepackage [english]{babel}
\usepackage [autostyle, english = american]{csquotes}
\MakeOuterQuote{"}

\begin{document}

\title{Emergence of long-range order in BaTiO$_{3}$ from local symmetry-breaking distortions}
\author{M. S. Senn}
\email{mark.senn@chem.ox.ac.uk}
\affiliation{Department of Chemistry, Inorganic Chemistry Laboratory, University of Oxford, South Parks Road, Oxford OX1 3QR, United Kingdom}
\author{D. A. Keen} 
\affiliation{ISIS, Rutherford Appleton Laboratory, Harwell Oxford, Didcot OX11 0QX, United Kingdom}
\author{T. C. A. Lucas}
\affiliation{School of Chemistry, University of Birmingham, Edgbaston, Birmingham, B15 2TT, United Kingdom}
\author{J. A. Hriljac}
\affiliation{School of Chemistry, University of Birmingham, Edgbaston, Birmingham, B15 2TT, United Kingdom}
\author{A. L. Goodwin}
\affiliation{Department of Chemistry, Inorganic Chemistry Laboratory, University of Oxford, South Parks Road, Oxford OX1 3QR, United Kingdom}

\date{\today}

\begin{abstract}

By using a symmetry motivated basis to evaluate local distortions against pair distribution function data (PDF), we show without prior bias, that the off-centre Ti displacements in the archetypal ferroelectric BaTiO$_{3}$ are zone centred and rhombohedral-like in nature across its known ferroelectric and paraelectric phases.  With our newly-gained insight we construct a simple Monte Carlo (MC) model which captures our main experimental findings and demonstrate how the rich crystallographic phase diagram of BaTiO$_{3}$ emerges from correlations of local symmetry-breaking distortions alone. Our results strongly support the order-disorder picture for these phase transitions, but can also be reconciled with the soft-mode theory of BaTiO$_{3}$ that is supported by some spectroscopic techniques.   

\end{abstract}


\maketitle

The phenomenological study of displacive phase transitions by Landau-Ginzburg theory in condensed matter science has been exceptionally fruitful~\cite{Cowley1980}. Its apparently close relationship to the theory of soft-mode phase transitions~\cite{Sollich1994,Dove1992} has meant that the expansion of the free energy of a system in terms of the correct order parameter can lead to a one-to-one correspondence with the energy of a phonon mode in the harmonic approximation.  Studies of global lattice and electronic instabilities benefit from the fact that global symmetry breaking (an order parameter) has been well classified in terms of irreducible representations (see ~\cite{Stokes2013} and references therein) of the parent symmetry space  which themselves have a correspondence with the eigenvectors (harmonic phonons) of the system under study.  Furthermore, their allowed couplings (or phonon-phonon scatterings) may be studied up to a given order by considering only those couplings which are invariant under the parent symmetry operators ~\cite{Hatch2003}. This, combined with the fact that global symmetry dictates both microscopic structure and macroscopic observables in these phase transitions means that the latter, which is often more conveniently measured, may be used as the order parameter for these phase transitions.  However, many macroscopic observables are a result of emergent phenomena driven by local ordering which combine in often counter-intuitive ways to produce the global symmetry of the structure and hence the local symmetry effectively controls the physical property ~\cite{Keen2015}.  Here the global symmetry is at odds with the local symmetry and a phenomenological model based around a macroscopic (or crystallographic) parameter will not lead to a valid physical insight into the phase transition.  Hence, microscopic studies of local symmetry-breaking are vital if physical understanding is to be gained into this class of phase transitions.          

In any crystallographic model there is, at least in principle, an infinitely large number of ways to locally break the average symmetry; for example, through thermal effects, distortions, or dislocations.  The difficulty is to coalesce these local degrees of freedom into an understandable form within a given length-scale and to evaluate their contribution systematically against data sensitive to local correlations such as XANES, EXAFS, PDF and diffuse scattering.  As no simple method exists for doing this, often our understanding of emergent phenomena in condensed phases is based on models building biased towards the macroscopic observables - whereas, ideally these properties should emerge spontaneously from our analysis of the local structure.   Here we present such a method and demonstrate its effectiveness at identifying the local order parameter in BaTiO$_{3}$.  Our results provide the first unbiased local model of this system and show how the crystallographic phase transitions can emerge naturally from the correlations of the local symmetry-breaking distortions.

\begin{figure}[t]
\includegraphics[width=8.5cm] {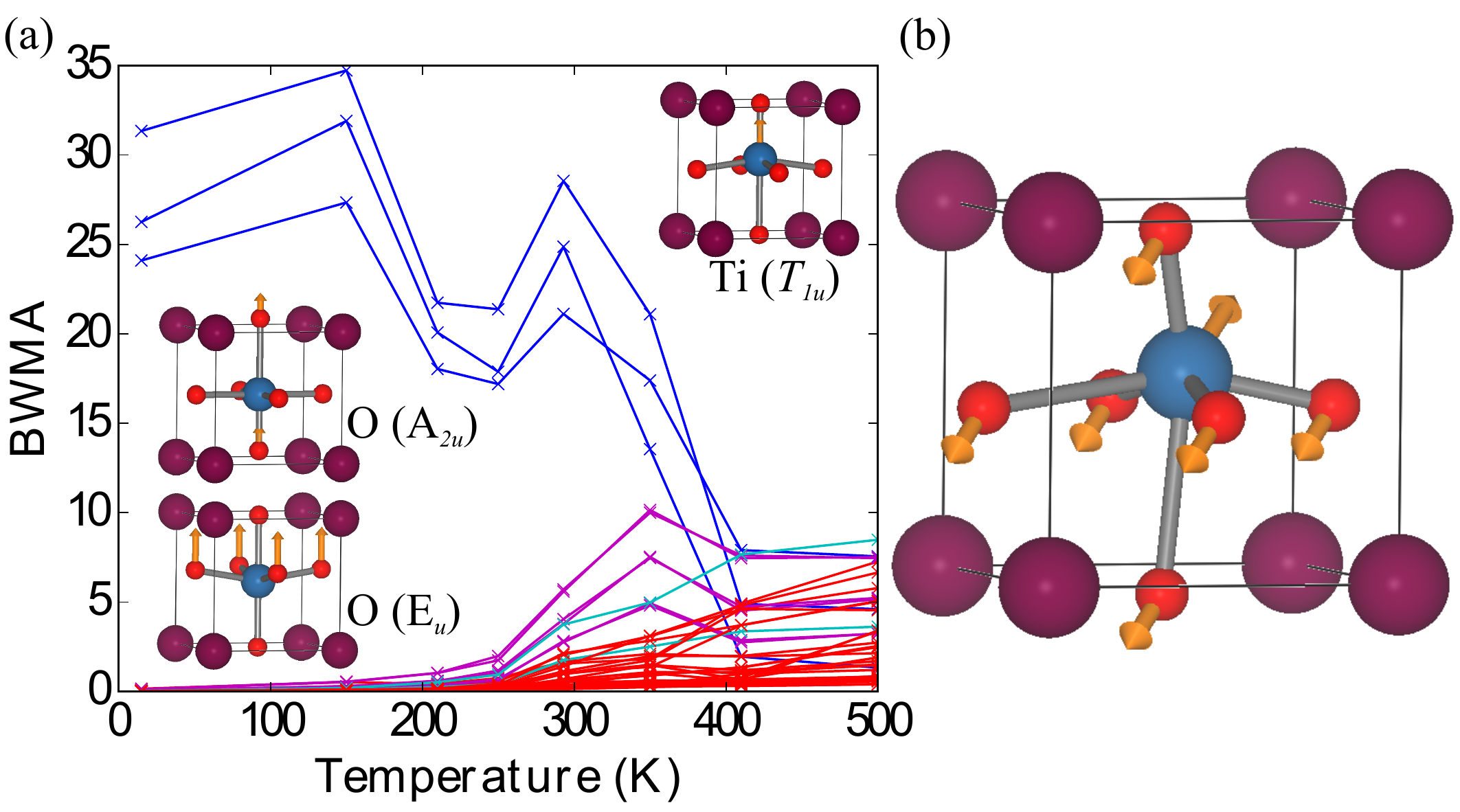}
\caption{\label{BWMA} (color online) $a$) BWMAs for all modes plotted against temperature.  Modes belonging to $\Gamma_{4}^{-}$, $X_{5}^{+}$, $M_{2}{-}$ and all other irreps. are colored blue, magenta, cyan and red respectively. Insets show symmetry-adapted displacements belonging to the $\Gamma_{4}^{-}$ irrep. which correspond to the three modes with the highest BWMA values and $b$) shows their combined + - - coupling (Ti($T_{1u}$), O($A_{2u}$) and O($E_{u}$) respectively) along the rhombohedral order parameter direction as observed locally in this study at all temperatures.}
\end{figure}

The ferroelectric properties of BaTiO$_{3}$ were first reported in the mid 1940s\cite{VonHippel1946}.  The origin of ferroelectricity in BaTiO$_{3}$ and the concomitant cubic to tetragonal structural distortion \cite{Megaw1947} was initially discussed in the framework of displacive phase transitions ~\cite{Cochran1959}.  However, the subsequent observations of orthorhombic and rhombohedral phase transitions at lower temperatures\cite{Kay1949,Rhodes1949} are inconsistent with second-order displacive phase transitions.   This anomaly in the symmetry relationship between low and high temperature ferroelectric phases together with the observation of diffuse x-ray scattering in all but the low temperature rhombohedral phase ~\cite{Comes1968} led to the development of the phenomenological order-disorder model ~\cite{Comes1970}.  This model has been supported by other techniques sensitive to local order such as PDF ~\cite{Kwei1995} and XANES/EXAFS methods ~\cite{Ravel1998}.   In this model, the incipient ferroelectric displacements of the Ti are taken to be rhombohedral-like, towards the faces of the TiO$_{6}$ octahedra (Fig. ~\ref{BWMA}($b$)), with long-range correlations in chains along $\left\langle 1 0 0 \right\rangle$, and disorder between chains giving rise to the average crystallographic symmetry.  The model explains the sheets of diffuse x-ray scattering observed in isostructural KNbO$_{3}$~\cite{Comes1970} and removes the requirement for the phase-transitions to proceed via group-subgroup relationships.  However, other diffuse scattering data have been analysed in terms of soft phonon modes ~\cite{Holma1995} and many spectroscopic techniques such as Raman ~\cite{Rimai1968,Vogt1982} and inelastic neutron ~\cite{Shirane1967,Shirane1970} scattering still appear to support a displacive phase transition model. Clearly consensus in the research community has yet to be reached.  We set out here to unambiguously identify the nature of the local displacements in the various reported crystallographic phases of BaTiO$_{3}$ and to reconcile long-range and short-range symmetry-breaking mechanisms in this important material.

Polycrystalline BaTiO$_{3}$ was prepared by conventional solid state synthesis. A 2 cm$^3$ sample was loaded into an 8 mm diameter cylindrical Vanadium can, and powder diffraction data were collected at 500, 410, 350, 293, 250, 210, 150 and 15 K on the instrument GEM ~\cite{Day2004} at the neutron spallation source ISIS, UK.   The maximum usable $Q$ was $40\,$\AA$^{-1}$ and counting times were 6-8 hrs. Rietveld refinements against the powder diffraction data were performed in the program Topas.  The total scattering data were then corrected for background, sample container, multiple scattering, absorption and inelasticity effects. using the program Gudrun ~\cite{McLain2012} to produce the total scattering structure factor $F(Q)$.  The sine Fourier transform of $F(Q)$ is the real-space pair distribution function (PDF).   As we are performing our fitting procedure in PDFFit~\cite{Farrow2007} we work here with $G^{\text{PDF}}(r)$ which is proportional to $D(r)$ and whose relationship with other parametrisations of the PDF is given in Ref.~\citenum{Keen2001}.

\begin{figure}[t]
\includegraphics[width=8.5cm] {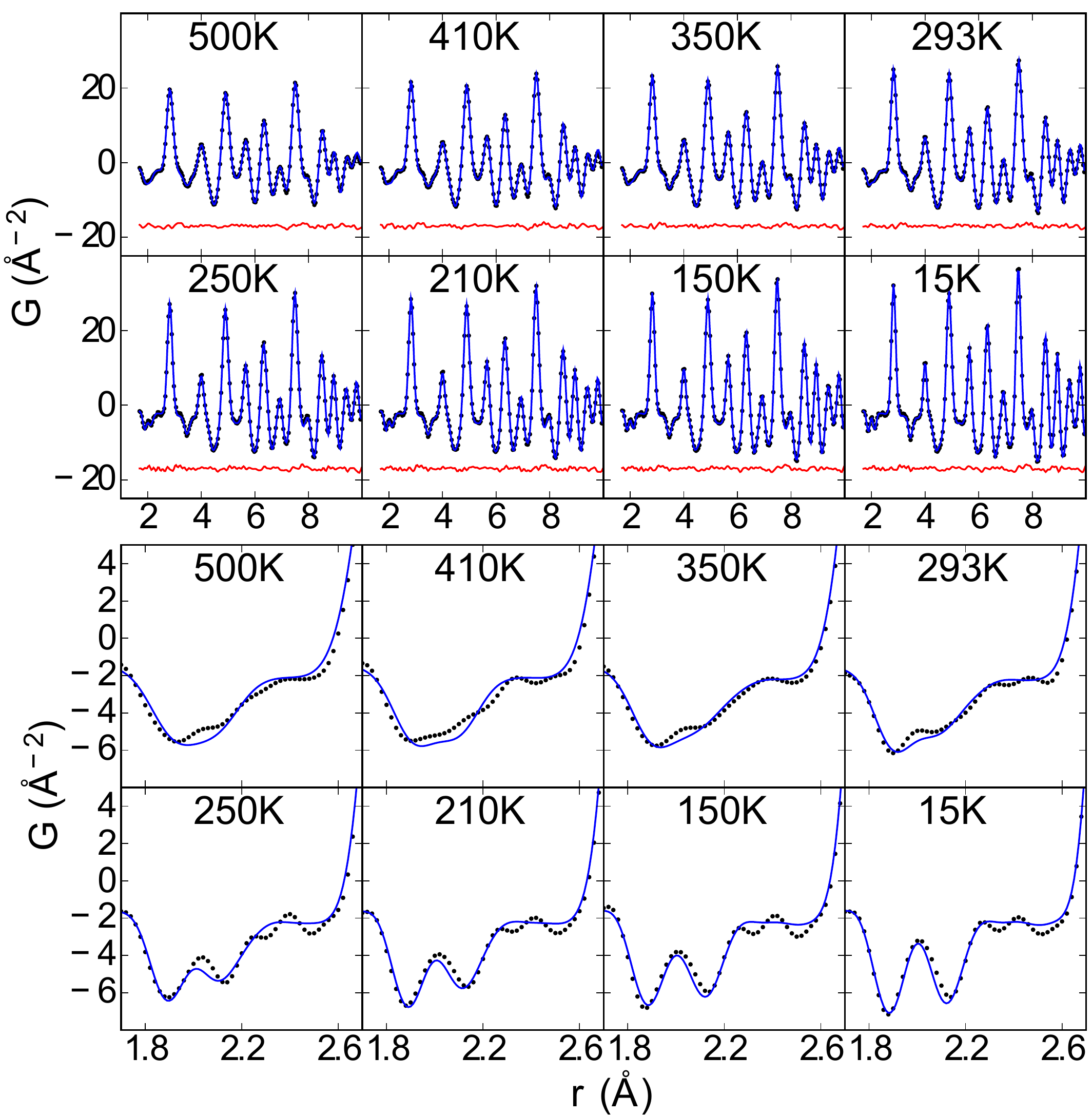}
\caption{\label{data_fits} (color online) Fits to G(r) at the various temperatures indicated using symmetry-adapted displacements belonging to to $\Gamma_{4}^{-}$ irrep. only. The fitting procedure is described in the text.}
\end{figure}

Although visual differences are evident in the PDFs as a function of temperature (Fig. ~\ref{data_fits}), some form of modelling is required to extract the information of interest - i.e. the local distortions of the atoms from their high symmetry positions. Two approaches exist here, one is a small box method fitting the PDF data whilst relying on a crystallographic unit cell akin to Rietveld fitting (e.g. ref. ~\cite{Billinge2004}) and the other is a big box method where a large supercell is constructed and reverse Monte Carlo (RMC) refinement is performed ~\cite{Playford2014}. The problem with conventional small box PDF analysis is that the parametrization used requires a priori assumption of the nature of the local distortions.  Equally, in RMC refinements, where the degrees of freedom are many, there are in principle a very large number of ways to interrogate the refined parameters and an exhaustive analysis of these correlations is impractical; bias is often introduced at this stage.  Recent work ~\cite{Neilson2015a} has sought to tackle this problem by decomposing RMC configurations in terms of their zone centre irreducible representations (irreps.). However, short range correlations beyond this length scale are lost in this analysis, and the modelling / refinement stage does not make use of the inherent orthogonality of symmetry adapted displacements. In contrast, the method we use here, involves expanding the degrees of freedom of the crystallographic unit cell up to a given supercell size in terms of both zone centre and zone boundary  irreps. of the "parent" space group.  The collection of symmetry-breaking displacements transforming as the same irrep. may be further decomposed into symmetry-adapted distortion modes (hereafter referred to simply as modes). There are now a number of web based tools which are able to perform this decomposition ~\cite{Campbell2006,PerezMato2010} and this parametrization has been used with much success in Rietveld refinement to systematically test for symmetry-breaking during phase transitions.~\cite{Campbell2007,Kerman2012}.  We extend this further here by utilizing the constraint language of PDFGui / PDFFit ~\cite{Farrow2007} such that we may refine mode amplitudes directly against PDF data.  Further details of the implementation are given as supplemental information (S.I.) and a comprehensive description will be published elsewhere.  

In Fig. ~\ref{data_fits} we show representative fits from our data analysis procedure corresponding to the global minimum observed for our refinements at each temperature.  We obtain very good fits to all the PDF data using this. Of course, the parametrisation of the $xyz$ degrees of freedom into modes does not change the overall goodness-of-fit but merely facilitates a systematic search of all parameter space. In the S.I. we also show that these fits are superior to those obtained with a big box RMC type model (noting that, although it has many more degrees of freedom, it is also constrained to simultaneously fit the Bragg profile) and that average models of the Rietveld analysis are inadequate at describing the PDF data at all but the lowest temperatures. 

We performed the analysis on a 2 $\times$ 2 $\times$ 2 $P1$ supercell of the $Pm\bar{3}m$ unit cell of paraelectric BaTiO$_{3}$, which can encompass all of the common symmetry lowering phase transitions observed in the perovskite family leading to 120 internal degrees of freedom, although, our analysis can easily be extended to incorporate displacements with longer period modulations.  We systematically tested the modes (by irrep.) against both Bragg and PDF data to emphasise when local and global symmetries agree or diverge. Each PDFgui or Rietveld refinement was initiated from random starting values 500 times at each temperature, so as to ensure that global minima were reached.  Modes with vanishingly-small refined amplitudes, or which produced poor fits, cannot be significant order parameters.  In Fig. ~\ref{BWMA} for each set of modes belonging to a irrep. tested in turn, we plot a weighted average of the refined mode magnitude from the 500 repeat refinements.  The weighting is such that large magnitudes from poorly fitting refinements corresponding to false minima have a vanishing contribution to the average through a Boltzmann distribution exp[(Rw$_{gobal}$-Rw)/$\sigma$).  We call this quantity the Boltzmann weighted mean amplitude (BWMA).  Rw$_{global}$ is the minimum value obtained in all refinements of the weighted R-factor and $\sigma$  is taken to be 0.1\%. The results do not change by considering $\sigma$ in a sensible interval and the value here is chosen such that two fits differing by this amount may only just be visually discriminated. Triply degenerate modes in the cubic symmetry corresponding to equivalent distortions in $x$, $y$ and $z$ directions are averaged together for these plots.  Three distinct types of behaviour of the BWMA are evident for different distortion modes: (1) have low values at 500 K rising to large values at lower temperatures; (2) have small values at 500 K dropping to near zero at low temperatures;(3)have consistently near zero values at all temperatures.  

Case (1) are clearly the relevant order parameters for the ferroelectric phase transitions. These modes all belong the $\Gamma_{4}^{-}$ irrep. and correspond to 9 degrees of freedom, three branches each for the Ti($T_{1u}$), O($A_{2u}$) and O($E_{u}$) modes.  Since modes belonging to different irreps. are necessarily orthogonal, there is no coupling (in the harmonic approximation), and we needn't test for inter-irrep. correlations, but instead can focus on intra-irrep. correlations, i.e. correlations within the basis describing the distortion space spanning this $\Gamma_{4}^{-}$  irrep.  Analysis of the three branches of each of these modes reveals that the order parameter always has rhombohedral symmetry from the local perspective, whereas our analysis of the Bragg profile in terms off global (crystallographic) symmetry, follows the well known cubic, tetragonal, orthorhombic and rhombohedral phase transitions described in the literature.  Full details of our analysis of the order parameter direction at all temperatures are given in in the SI.  Our results are consistent with recent RMC refinements on the tetragonal phase of BaTiO$_{3}$ which find a Ti distribution in agreement with a picture of local rhombohedral-like distortions ~\cite{Levin2014}, however, the clear off-centre nature of the Ti distributions that we observe, does not appear to be evident in that work.  Case (2) demonstrates our sensitivity to soft phonon modes.  These modes belong to irrep. $X_{5}^{+}$ or $M_{2}{-}$, are soft eigenvectors of the system \cite{Ghosez1997b}, and are on the same line in the phonon dispersion curves as the zone centred $\Gamma_{4}^{-}$ ferroelectric instability, such that they both correspond to locally the same off-centre distortions, but are antiferroelectric in nature.  Case (3) shows that a large number of modes do not have a significant contribution to describing the local symmetry breaking distortions in BaTiO$_{3}$.

\begin{figure*}
\includegraphics[width=\textwidth]{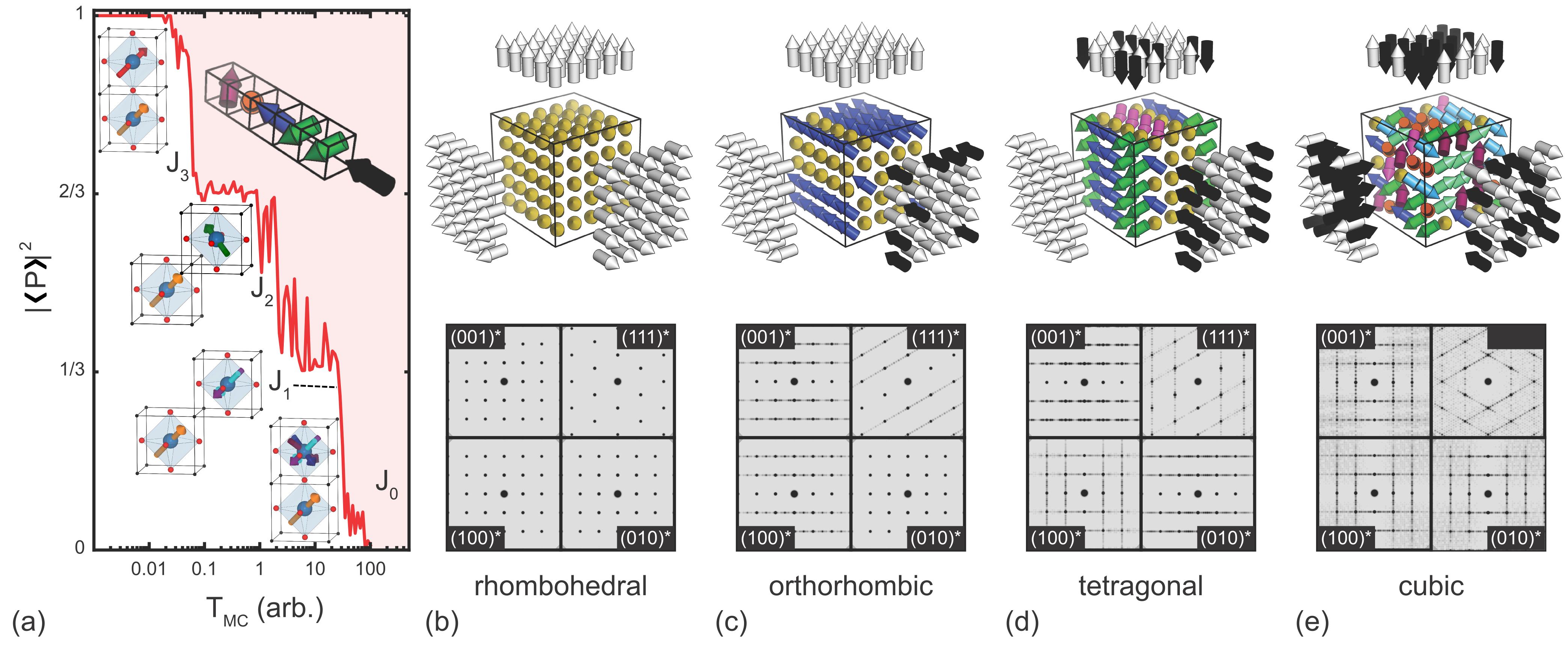}
\caption{\label{MC} (color online) $a$) MC simulation reproducing the tetragonal, orthorhombic and rhombohedral phase transitions, with the square of the average polarization serving as an order parameter. Insets below the polarization line show a graphical visualization for the relevant interactions penalized by our Hamiltonian, shown at the temperatures their energy scales become important in the MC simulation, and above the line, a representation of the chain like correlations present in the cubic phase (J$_{0}$, hardwired into our simulations) showing how the projection of the polarization along the direction of a chain is always preserved (black arrow). $b$)-$e$) a portion of an MC configuration for each phase, along with their polar vector projections along $\left\langle 1 0 0 \right\rangle$ directions, and their calculated diffuse scattering in the planes indicated.  The polar vector projections (white and black arrows) represent columns of Ti displacements all with a common component along the chain, and it is this inter-chain disorder which gives rise to diffuse planes of scattering in reciprocal space.  Inter-chain disorder in the cubic phase in all three $\left\langle 1 0 0 \right\rangle$ directions, leads to three sets of intersecting orthogonal diffuse scattering planes in reciprocal space, and each successive phase transition represent inter-chain orderings occurring along successive $\left\langle 1 0 0 \right\rangle$ directions.}
\end{figure*}

The origin of the local order parameter is a strong desire for Ti to undergo a second order Jahn-Teller distortion, presumably driven by a hybridization of the oxygen 2$p$ orbitals with the Ti 3$d$ orbitals~ \cite{Cohen1992} by shortening three bonds simultaneously, i.e. a local rhombohedral like distortion.  The incipient rhombohedral distortions are necessarily split by global tetragonal and orthorhombic lattice distortions at particular temperatures, and the question now arises, can these lattice distortions themselves be a result of correlations of the local symmetry-breaking distortions alone?  Or put another way, can this order-disorder model provide a complete description of the observed phase transitions in BaTiO$_{3}$?   

We construct a simple Hamiltonian to illustrate how the long-range crystallographic symmetry can emerge from local correlations of the rhombohedral-like Ti displacements.  We make no claim that our model represents the true physical Hamiltonian of the system, but instead use the model to show that local coupling of Ti displacements can reproduce the observed series of crystallographic phase transitions and is consistent with previously reported diffuse scattering ~\cite{Comes1970}.  The Hamiltonian that we use in this MC simulation is of the form
\begin{equation*}
\mathcal{H}=J_1\sum_{\langle\langle i,j\rangle\rangle}\delta\left[\frac{\sqrt{3}}{2\sqrt{2}}(\mathbf S_i^\parallel-\mathbf S_j^\parallel)-\hat{\mathbf r}_{ij}\right]
\end{equation*} 
\begin{equation*}
+J_2\sum_{\langle\langle\langle i,j\rangle\rangle\rangle}\left[\delta(\mathbf S_i-\hat{\mathbf r}_{ij})\times\delta\left(\mathbf S_j\cdot\hat{\mathbf r}_{ij}+\frac{1}{3}\right)\right] + J_3\sum_{\langle i,j\rangle}\mathbf S_i\cdot\mathbf S_j
\end{equation*}

Here $\mathbf S_i$ denotes the normalised displacement vector of Ti site $i$ and $\hat{\mathbf r}_{ij}$ the normalised vector joining sites $i$ and $j$. The coupling parameters are of the form $J_1,J_2>0$, $J_3<0$ and $|J_3|<|J_2|<|J_1|$. The notation $\langle i,j\rangle$ denotes a sum over nearest-neighbour sites $i,j$; double and triple angle brackets denote sums over second and third nearest neighbours, respectively.  The three terms in this toy Hamiltonian penalise dipole misalignment in first-, second-, and third-nearest neighbour sites as shown in the inset of Fig. ~\ref{MC} ($a$).  We use this Hamiltonian as the cost function for a MC simulation on a 10 $\times$ 10 $\times$ 10 supercell in which all dipoles are hard-wired to have a common projection along any one chain when viewed down one of the $\left\langle 1 0 0 \right\rangle$ directions (i.e. we assume a J$_{0}$ term with high energy in keeping with Ref. ~\citenum{Comes1970} and as illustrated in the inset of Fig. ~\ref{MC} ($a$)).   The results of this simulation (Fig. ~\ref{MC} ($a$)) show three distinct phase transitions as evidenced by discontinuities in macroscopic polarization.  In Fig. ~\ref{MC} ($b$)-($e$) we illustrate portions of representative configurations of the correlated disorder observed in the MC simulation, along with their calculated diffuse scattering.  Both our calculated diffuse scattering and observed chain like configurations of off-centre Ti displacements are consistent with that published by Comes et al. \cite{Comes1968,Comes1970}.

These MC simulations show how the observed sequence of phase transitions can arise from local correlations of nearest-neighbour type interactions. They strongly support the order-disorder picture of BaTiO$_{3}$ and show how local correlations alone are capable of giving rise to the rich phase diagram observed for this material. Importantly, this order-disorder picture is not necessarily contradictory to spectroscopic results which evidence a soft-mode picture ~\cite{Rimai1968,Vogt1982,Shirane1967,Shirane1970}.  When our high temperature cubic phase is viewed down any of its 4-fold axes, the correlated disorder of the rhomohedral-like Ti displacements clearly have average tetragonal symmetry for any one given chain.  There are 6 symmetry equivalent tetragonal-like chains, in which the average Ti displacements sit along one of the $\left\langle 1 0 0 \right\rangle$ directions.  A spontaneous symmetry-breaking event in which the correlation length of one of these chains tends to infinity could be viewed as a pseudo-tetragonal soft mode phase transition.  

Our results are the first unbiased determination of local symmetry in BaTiO$_{3}$ across its known ferroelectric and paraelectric phases.  They show that local displacements of the Ti atoms are zone centered and rhombohedral-like at all temperatures.  The fact that a simple Hamiltonian, which considers only local interactions, can reproduce the rich phase diagram in an MC simulation gives strong support to the order-disorder picture for BaTiO$_{3}$ and demonstrates how global symmetry may emerge from local symmetry-breaking distortions.  This fundamental insight highlights the importance of our methodology for determining local symmetry in order-disorder phase transitions and demonstrates its power when it is coupled to MC simulations. We note that recent developments in speeding up the simulation of PDF data by orders of magnitude ~\cite{Coelho2015} will allow the exploration of ever more complicated systems exhibiting order-disorder phase transitions, and hope that our work will stimulate research in this spirit into a host of other systems exhibiting emergent phenomena.  

We would like to acknowledge STFC for awarding us beamtime to conduct these experiments. MSS is grateful to the Royal Commission for the Exhibition of 1851 for a fellowship, ALG for funding from the ERC (grant no. 279705) and TCAL and JAH for studentship funding from Diamond Light Source.


\begin{thebibliography}{35}
\expandafter\ifx\csname natexlab\endcsname\relax\def\natexlab#1{#1}\fi
\expandafter\ifx\csname bibnamefont\endcsname\relax
  \def\bibnamefont#1{#1}\fi
\expandafter\ifx\csname bibfnamefont\endcsname\relax
  \def\bibfnamefont#1{#1}\fi
\expandafter\ifx\csname citenamefont\endcsname\relax
  \def\citenamefont#1{#1}\fi
\expandafter\ifx\csname url\endcsname\relax
  \def\url#1{\texttt{#1}}\fi
\expandafter\ifx\csname urlprefix\endcsname\relax\def\urlprefix{URL }\fi
\providecommand{\bibinfo}[2]{#2}
\providecommand{\eprint}[2][]{\url{#2}}



\bibitem [{\citenamefont {Cowley}(1980)}]{Cowley1980}%
  \BibitemOpen
  \bibfield  {author} {\bibinfo {author} {\bibfnamefont {R.~A.}\ \bibnamefont
  {Cowley}},\ }\href {\doibase 10.1080/00018738000101346} {\bibfield  {journal}
  {\bibinfo  {journal} {Adv. Phys.}\ }\textbf {\bibinfo {volume} {29}},\
  \bibinfo {pages} {1} (\bibinfo {year} {1980})}\BibitemShut {NoStop}%
\bibitem [{\citenamefont {Sollich}\ \emph {et~al.}(1994)\citenamefont
  {Sollich}, \citenamefont {Heine},\ and\ \citenamefont {Dove}}]{Sollich1994}%
  \BibitemOpen
  \bibfield  {author} {\bibinfo {author} {\bibfnamefont {P.}~\bibnamefont
  {Sollich}}, \bibinfo {author} {\bibfnamefont {V.}~\bibnamefont {Heine}}, \
  and\ \bibinfo {author} {\bibfnamefont {M.~T.}\ \bibnamefont {Dove}},\
  }\href {} {\bibfield  {journal} {\bibinfo  {journal} {J.
  Physics-Condensed Matter}\ }\textbf {\bibinfo {volume} {6}},\ \bibinfo
  {pages} {3171} (\bibinfo {year} {1994})}\BibitemShut {NoStop}%
\bibitem [{\citenamefont {Dove}\ \emph {et~al.}(1992)\citenamefont {Dove},
  \citenamefont {Giddy},\ and\ \citenamefont {Heine}}]{Dove1992}%
  \BibitemOpen
  \bibfield  {author} {\bibinfo {author} {\bibfnamefont {M.~T.}\ \bibnamefont
  {Dove}}, \bibinfo {author} {\bibfnamefont {A.~P.}\ \bibnamefont {Giddy}}, \
  and\ \bibinfo {author} {\bibfnamefont {V.}~\bibnamefont {Heine}},\ }\href
  {\doibase 10.1080/00150199208016064} {\bibfield  {journal} {\bibinfo
  {journal} {Ferroelectrics}\ }\textbf {\bibinfo {volume} {136}},\ \bibinfo
  {pages} {33} (\bibinfo {year} {1992})}\BibitemShut {NoStop}%
\bibitem [{\citenamefont {Stokes}\ \emph {et~al.}(2013)\citenamefont {Stokes},
  \citenamefont {Campbell},\ and\ \citenamefont {Cordes}}]{Stokes2013}%
  \BibitemOpen
  \bibfield  {author} {\bibinfo {author} {\bibfnamefont {H.~T.}\ \bibnamefont
  {Stokes}}, \bibinfo {author} {\bibfnamefont {B.~J.}\ \bibnamefont
  {Campbell}}, \ and\ \bibinfo {author} {\bibfnamefont {R.}~\bibnamefont
  {Cordes}},\ }\href {\doibase 10.1107/S0108767313007538} {\bibfield  {journal}
  {\bibinfo  {journal} {Acta Crystallogr. Sect. A Found. Crystallogr.}\
  }\textbf {\bibinfo {volume} {69}},\ \bibinfo {pages} {388} (\bibinfo {year}
  {2013})}\BibitemShut {NoStop}%
\bibitem [{\citenamefont {Hatch}\ and\ \citenamefont
  {Stokes}(2003)}]{Hatch2003}%
  \BibitemOpen
  \bibfield  {author} {\bibinfo {author} {\bibfnamefont {D.~M.}\ \bibnamefont
  {Hatch}}\ and\ \bibinfo {author} {\bibfnamefont {H.~T.}\ \bibnamefont
  {Stokes}},\ }\href {\doibase 10.1107/S0021889803005946} {\bibfield  {journal}
  {\bibinfo  {journal} {J. Appl. Crystallogr.}\ }\textbf {\bibinfo {volume}
  {36}},\ \bibinfo {pages} {951} (\bibinfo {year} {2003})}\BibitemShut
  {NoStop}%
\bibitem [{\citenamefont {Keen}\ and\ \citenamefont
  {Goodwin}(2015)}]{Keen2015}%
  \BibitemOpen
  \bibfield  {author} {\bibinfo {author} {\bibfnamefont {D.~A.}\ \bibnamefont
  {Keen}}\ and\ \bibinfo {author} {\bibfnamefont {A.~L.}\ \bibnamefont
  {Goodwin}},\ }\href {\doibase 10.1038/nature14453} {\bibfield  {journal}
  {\bibinfo  {journal} {Nature}\ }\textbf {\bibinfo {volume} {521}},\ \bibinfo
  {pages} {303} (\bibinfo {year} {2015})}\BibitemShut {NoStop}%
\bibitem [{\citenamefont {{Von Hippel}}\ \emph {et~al.}(1946)\citenamefont
  {{Von Hippel}}, \citenamefont {Breckenridge}, \citenamefont {Chesley},\ and\
  \citenamefont {Tisza}}]{VonHippel1946}%
  \BibitemOpen
  \bibfield  {author} {\bibinfo {author} {\bibfnamefont {A.}~\bibnamefont {{Von
  Hippel}}}, \bibinfo {author} {\bibfnamefont {R.~G.}\ \bibnamefont
  {Breckenridge}}, \bibinfo {author} {\bibfnamefont {F.~G.}\ \bibnamefont
  {Chesley}}, \ and\ \bibinfo {author} {\bibfnamefont {L.}~\bibnamefont
  {Tisza}},\ }\href {\doibase 10.1021/ie50443a009} {\bibfield  {journal}
  {\bibinfo  {journal} {Ind. Eng. Chem.}\ }\textbf {\bibinfo {volume} {38}},\
  \bibinfo {pages} {1097} (\bibinfo {year} {1946})}\BibitemShut {NoStop}%
\bibitem [{\citenamefont {Megaw}(1947)}]{Megaw1947}%
  \BibitemOpen
  \bibfield  {author} {\bibinfo {author} {\bibfnamefont {H.~D.}\ \bibnamefont
  {Megaw}},\ }\href {\doibase 10.1098/rspa.1947.0038} {\bibfield  {journal}
  {\bibinfo  {journal} {Proc. R. Soc. A Math. Phys. Eng. Sci.}\ }\textbf
  {\bibinfo {volume} {189}},\ \bibinfo {pages} {261} (\bibinfo {year}
  {1947})}\BibitemShut {NoStop}%
\bibitem [{\citenamefont {Cochran}(1960)}]{Cochran1959}%
  \BibitemOpen
  \bibfield  {author} {\bibinfo {author} {\bibfnamefont {W.}~\bibnamefont
  {Cochran}},\ }\href {\doibase 10.1103/PhysRevLett.3.412} {\bibfield
  {journal} {\bibinfo  {journal} {Adv. Condens. Matter Phys.}\ }\textbf
  {\bibinfo {volume} {9}},\ \bibinfo {pages} {387} (\bibinfo {year}
  {1960})}\BibitemShut {NoStop}%
\bibitem [{\citenamefont {Kay}\ and\ \citenamefont {Vousden}(1949)}]{Kay1949}%
  \BibitemOpen
  \bibfield  {author} {\bibinfo {author} {\bibfnamefont {H.~F.}\ \bibnamefont
  {Kay}}\ and\ \bibinfo {author} {\bibfnamefont {P.}~\bibnamefont {Vousden}},\
  }\href {\doibase 10.1080/14786444908561371} {\bibfield  {journal} {\bibinfo
  {journal} {Philos. Mag. Ser. 7}\ }\textbf {\bibinfo {volume} {40}},\ \bibinfo
  {pages} {1019} (\bibinfo {year} {1949})}\BibitemShut {NoStop}%
\bibitem [{\citenamefont {Rhodes}(1949)}]{Rhodes1949}%
  \BibitemOpen
  \bibfield  {author} {\bibinfo {author} {\bibfnamefont {R.~G.}\ \bibnamefont
  {Rhodes}},\ }\href {\doibase 10.1107/S0365110X49001077} {\bibfield  {journal}
  {\bibinfo  {journal} {Acta Crystallogr.}\ }\textbf {\bibinfo {volume} {2}},\
  \bibinfo {pages} {417} (\bibinfo {year} {1949})}\BibitemShut {NoStop}%
\bibitem [{\citenamefont {Com{\`{e}}s}\ \emph {et~al.}(1968)\citenamefont
  {Com{\`{e}}s}, \citenamefont {Lambert},\ and\ \citenamefont
  {Guinier}}]{Comes1968}%
  \BibitemOpen
  \bibfield  {author} {\bibinfo {author} {\bibfnamefont {R.}~\bibnamefont
  {Com{\`{e}}s}}, \bibinfo {author} {\bibfnamefont {M.}~\bibnamefont
  {Lambert}}, \ and\ \bibinfo {author} {\bibfnamefont {A.}~\bibnamefont
  {Guinier}},\ }\href {\doibase 10.1016/0038-1098(68)90571-1} {\bibfield
  {journal} {\bibinfo  {journal} {Solid State Commun.}\ }\textbf {\bibinfo
  {volume} {6}},\ \bibinfo {pages} {715} (\bibinfo {year} {1968})}\BibitemShut
  {NoStop}%
\bibitem [{\citenamefont {Com{\`{e}}s}\ \emph {et~al.}(1970)\citenamefont
  {Com{\`{e}}s}, \citenamefont {Lambert},\ and\ \citenamefont
  {Guinier}}]{Comes1970}%
  \BibitemOpen
  \bibfield  {author} {\bibinfo {author} {\bibfnamefont {R.}~\bibnamefont
  {Com{\`{e}}s}}, \bibinfo {author} {\bibfnamefont {M.}~\bibnamefont
  {Lambert}}, \ and\ \bibinfo {author} {\bibfnamefont {A.}~\bibnamefont
  {Guinier}},\ }\href {\doibase 10.1107/S056773947000061X} {\bibfield
  {journal} {\bibinfo  {journal} {Acta Crystallogr. Sect. A Cryst. Physics,
  Diffraction, Theor. Gen. Crystallogr.}\ }\textbf {\bibinfo {volume} {26}},\
  \bibinfo {pages} {244} (\bibinfo {year} {1970})}\BibitemShut {NoStop}%
\bibitem [{\citenamefont {Kwei}\ \emph {et~al.}(1995)\citenamefont {Kwei},
  \citenamefont {Billinge}, \citenamefont {Cheong},\ and\ \citenamefont
  {Saxton}}]{Kwei1995}%
  \BibitemOpen
  \bibfield  {author} {\bibinfo {author} {\bibfnamefont {G.~H.}\ \bibnamefont
  {Kwei}}, \bibinfo {author} {\bibfnamefont {S.~J.~L.}\ \bibnamefont
  {Billinge}}, \bibinfo {author} {\bibfnamefont {S.-W.}\ \bibnamefont
  {Cheong}}, \ and\ \bibinfo {author} {\bibfnamefont {J.~G.}\ \bibnamefont
  {Saxton}},\ }\href {\doibase 10.1080/00150199508221830} {\bibfield  {journal}
  {\bibinfo  {journal} {Ferroelectrics}\ }\textbf {\bibinfo {volume} {164}},\
  \bibinfo {pages} {57} (\bibinfo {year} {1995})}\BibitemShut {NoStop}%
\bibitem [{\citenamefont {Ravel}\ \emph {et~al.}(1998)\citenamefont {Ravel},
  \citenamefont {Stern}, \citenamefont {Vedrinskii},\ and\ \citenamefont
  {Kraizman}}]{Ravel1998}%
  \BibitemOpen
  \bibfield  {author} {\bibinfo {author} {\bibfnamefont {B.}~\bibnamefont
  {Ravel}}, \bibinfo {author} {\bibfnamefont {E.~A.}\ \bibnamefont {Stern}},
  \bibinfo {author} {\bibfnamefont {R.~I.}\ \bibnamefont {Vedrinskii}}, \ and\
  \bibinfo {author} {\bibfnamefont {V.}~\bibnamefont {Kraizman}},\ }\href
  {\doibase 10.1080/00150199808009173} {\bibfield  {journal} {\bibinfo
  {journal} {Ferroelectrics}\ }\textbf {\bibinfo {volume} {206}},\ \bibinfo
  {pages} {407} (\bibinfo {year} {1998})}\BibitemShut {NoStop}%
\bibitem [{\citenamefont {Holma}\ \emph {et~al.}(1995)\citenamefont {Holma},
  \citenamefont {Takesue},\ and\ \citenamefont {Chen}}]{Holma1995}%
  \BibitemOpen
  \bibfield  {author} {\bibinfo {author} {\bibfnamefont {M.}~\bibnamefont
  {Holma}}, \bibinfo {author} {\bibfnamefont {N.~N.}\ \bibnamefont {Takesue}},
  \ and\ \bibinfo {author} {\bibfnamefont {H.}~\bibnamefont {Chen}},\ }\href
  {\doibase 10.1080/00150199508221847} {\bibfield  {journal} {\bibinfo
  {journal} {Ferroelectrics}\ }\textbf {\bibinfo {volume} {164}},\ \bibinfo
  {pages} {237} (\bibinfo {year} {1995})}\BibitemShut {NoStop}%
\bibitem [{\citenamefont {Rimai}\ \emph {et~al.}(1968)\citenamefont {Rimai},
  \citenamefont {Parsons},\ and\ \citenamefont {Hickmott}}]{Rimai1968}%
  \BibitemOpen
  \bibfield  {author} {\bibinfo {author} {\bibfnamefont {L.}~\bibnamefont
  {Rimai}}, \bibinfo {author} {\bibfnamefont {J.~L.}\ \bibnamefont {Parsons}},
  \ and\ \bibinfo {author} {\bibfnamefont {J.~T.}\ \bibnamefont {Hickmott}},\
  }\href {} {\bibfield  {journal} {\bibinfo  {journal} {Phys. Rev.}\
  }\textbf {\bibinfo {volume} {168}},\ \bibinfo {pages} {623} (\bibinfo {year}
  {1968})}\BibitemShut {NoStop}%
\bibitem [{\citenamefont {Vogt}\ \emph {et~al.}(1982)\citenamefont {Vogt},
  \citenamefont {Sanjurjo},\ and\ \citenamefont {Rossbroich}}]{Vogt1982}%
  \BibitemOpen
  \bibfield  {author} {\bibinfo {author} {\bibfnamefont {H.}~\bibnamefont
  {Vogt}}, \bibinfo {author} {\bibfnamefont {J.}~\bibnamefont {Sanjurjo}}, \
  and\ \bibinfo {author} {\bibfnamefont {G.}~\bibnamefont {Rossbroich}},\
  }\href {\doibase 10.1103/PhysRevB.26.5904} {\bibfield  {journal} {\bibinfo
  {journal} {Phys. Rev. B}\ }\textbf {\bibinfo {volume} {26}},\ \bibinfo
  {pages} {5904} (\bibinfo {year} {1982})}\BibitemShut {NoStop}%
\bibitem [{\citenamefont {Shirane}\ \emph {et~al.}(1967)\citenamefont
  {Shirane}, \citenamefont {Frazer}, \citenamefont {Minkiewicz}, \citenamefont
  {Leake},\ and\ \citenamefont {Linz}}]{Shirane1967}%
  \BibitemOpen
  \bibfield  {author} {\bibinfo {author} {\bibfnamefont {G.}~\bibnamefont
  {Shirane}}, \bibinfo {author} {\bibfnamefont {B.}~\bibnamefont {Frazer}},
  \bibinfo {author} {\bibfnamefont {V.}~\bibnamefont {Minkiewicz}}, \bibinfo
  {author} {\bibfnamefont {J.}~\bibnamefont {Leake}}, \ and\ \bibinfo {author}
  {\bibfnamefont {A.}~\bibnamefont {Linz}},\ }\href {\doibase
  10.1103/PhysRevLett.19.234} {\bibfield  {journal} {\bibinfo  {journal} {Phys.
  Rev. Lett.}\ }\textbf {\bibinfo {volume} {19}},\ \bibinfo {pages} {234}
  (\bibinfo {year} {1967})}\BibitemShut {NoStop}%
\bibitem [{\citenamefont {Shirane}\ \emph {et~al.}(1970)\citenamefont
  {Shirane}, \citenamefont {Axe},\ and\ \citenamefont {Harada}}]{Shirane1970}%
  \BibitemOpen
  \bibfield  {author} {\bibinfo {author} {\bibfnamefont {G.}~\bibnamefont
  {Shirane}}, \bibinfo {author} {\bibfnamefont {J.~D.}\ \bibnamefont {Axe}}, \
  and\ \bibinfo {author} {\bibfnamefont {J.}~\bibnamefont {Harada}},\
  }\href {} {\bibfield  {journal} {\bibinfo  {journal} {Phys. Rev. B}\
  }\textbf {\bibinfo {volume} {2}},\ \bibinfo {pages} {3651} (\bibinfo {year}
  {1970})}\BibitemShut {NoStop}%
\bibitem [{\citenamefont {Day}\ \emph {et~al.}(2004)\citenamefont {Day},
  \citenamefont {Enderby}, \citenamefont {Williams}, \citenamefont {Chapon},
  \citenamefont {Hannon}, \citenamefont {Radaelli},\ and\ \citenamefont
  {Soper}}]{Day2004}%
  \BibitemOpen
  \bibfield  {author} {\bibinfo {author} {\bibfnamefont {P.}~\bibnamefont
  {Day}}, \bibinfo {author} {\bibfnamefont {J.}~\bibnamefont {Enderby}},
  \bibinfo {author} {\bibfnamefont {W.}~\bibnamefont {Williams}}, \bibinfo
  {author} {\bibfnamefont {L.}~\bibnamefont {Chapon}}, \bibinfo {author}
  {\bibfnamefont {A.~C.}\ \bibnamefont {Hannon}}, \bibinfo {author}
  {\bibfnamefont {P.}~\bibnamefont {Radaelli}}, \ and\ \bibinfo {author}
  {\bibfnamefont {A.~K.}\ \bibnamefont {Soper}},\ }\href {\doibase
  10.1080/00323910490970564} {\bibfield  {journal} {\bibinfo  {journal}
  {Neutron News}\ }\textbf {\bibinfo {volume} {15}},\ \bibinfo {pages} {19}
  (\bibinfo {year} {2004})}\BibitemShut {NoStop}%
\bibitem [{\citenamefont {McLain}\ \emph {et~al.}(2012)\citenamefont {McLain},
  \citenamefont {Bowron}, \citenamefont {Hannon},\ and\ \citenamefont
  {Soper}}]{McLain2012}%
  \BibitemOpen
  \bibfield  {author} {\bibinfo {author} {\bibfnamefont {S.~E.}\ \bibnamefont
  {McLain}}, \bibinfo {author} {\bibfnamefont {D.~T.}\ \bibnamefont {Bowron}},
  \bibinfo {author} {\bibfnamefont {A.~C.}\ \bibnamefont {Hannon}}, \ and\
  \bibinfo {author} {\bibfnamefont {A.~K.}\ \bibnamefont {Soper}},\ }\href
  {} {\enquote {\bibinfo {title} {{GUDRUN, a computer program developed for
  analysis of neutron diffraction data, Chilton: ISIS Facility, Rutherford
  Appleton Laboratory}},}\ } (\bibinfo {year} {2012})\BibitemShut {NoStop}%
\bibitem [{\citenamefont {Farrow}\ \emph {et~al.}(2007)\citenamefont {Farrow},
  \citenamefont {Juhas}, \citenamefont {Liu}, \citenamefont {Bryndin},
  \citenamefont {Bo{\v{z}}in}, \citenamefont {Bloch}, \citenamefont {Proffen},\
  and\ \citenamefont {Billinge}}]{Farrow2007}%
  \BibitemOpen
  \bibfield  {author} {\bibinfo {author} {\bibfnamefont {C.~L.}\ \bibnamefont
  {Farrow}}, \bibinfo {author} {\bibfnamefont {P.}~\bibnamefont {Juhas}},
  \bibinfo {author} {\bibfnamefont {J.~W.}\ \bibnamefont {Liu}}, \bibinfo
  {author} {\bibfnamefont {D.}~\bibnamefont {Bryndin}}, \bibinfo {author}
  {\bibfnamefont {E.~S.}\ \bibnamefont {Bo{\v{z}}in}}, \bibinfo {author}
  {\bibfnamefont {J.}~\bibnamefont {Bloch}}, \bibinfo {author} {\bibfnamefont
  {T.}~\bibnamefont {Proffen}}, \ and\ \bibinfo {author} {\bibfnamefont
  {S.~J.~L.}\ \bibnamefont {Billinge}},\ }\href {\doibase
  10.1088/0953-8984/19/33/335219} {\bibfield  {journal} {\bibinfo  {journal}
  {J. Phys. Condens. Matter}\ }\textbf {\bibinfo {volume} {19}},\ \bibinfo
  {pages} {335219} (\bibinfo {year} {2007})}\BibitemShut {NoStop}%
\bibitem [{\citenamefont {Keen}(2001)}]{Keen2001}%
  \BibitemOpen
  \bibfield  {author} {\bibinfo {author} {\bibfnamefont {D.~A.}\ \bibnamefont
  {Keen}},\ }\href {\doibase 10.1107/S0021889800019993} {\bibfield  {journal}
  {\bibinfo  {journal} {J. Appl. Crystallogr.}\ }\textbf {\bibinfo {volume}
  {34}},\ \bibinfo {pages} {172} (\bibinfo {year} {2001})}\BibitemShut
  {NoStop}%
\bibitem [{\citenamefont {Billinge}(2004)}]{Billinge2004}%
  \BibitemOpen
  \bibfield  {author} {\bibinfo {author} {\bibfnamefont {S.~J.~L.}\
  \bibnamefont {Billinge}},\ }\href {\doibase 10.1524/zkri.219.3.117.29094}
  {\bibfield  {journal} {\bibinfo  {journal} {Zeitschrift f{\"{u}}r Krist.}\
  }\textbf {\bibinfo {volume} {219}},\ \bibinfo {pages} {117} (\bibinfo {year}
  {2004})}\BibitemShut {NoStop}%
\bibitem [{\citenamefont {Playford}\ \emph {et~al.}(2014)\citenamefont
  {Playford}, \citenamefont {Owen}, \citenamefont {Levin},\ and\ \citenamefont
  {Tucker}}]{Playford2014}%
  \BibitemOpen
  \bibfield  {author} {\bibinfo {author} {\bibfnamefont {H.~Y.}\ \bibnamefont
  {Playford}}, \bibinfo {author} {\bibfnamefont {L.~R.}\ \bibnamefont {Owen}},
  \bibinfo {author} {\bibfnamefont {I.}~\bibnamefont {Levin}}, \ and\ \bibinfo
  {author} {\bibfnamefont {M.~G.}\ \bibnamefont {Tucker}},\ }\href {\doibase
  10.1146/annurev-matsci-071312-121712} {\bibfield  {journal} {\bibinfo
  {journal} {Annu. Rev. Mater. Res.}\ }\textbf {\bibinfo {volume} {44}},\
  \bibinfo {pages} {429} (\bibinfo {year} {2014})}\BibitemShut {NoStop}%
\bibitem [{\citenamefont {Neilson}\ and\ \citenamefont
  {McQueen}(2015)}]{Neilson2015a}%
  \BibitemOpen
  \bibfield  {author} {\bibinfo {author} {\bibfnamefont {J.~R.}\ \bibnamefont
  {Neilson}}\ and\ \bibinfo {author} {\bibfnamefont {T.~M.}\ \bibnamefont
  {McQueen}},\ }\href {\doibase 10.1107/S1600576715016404} {\bibfield
  {journal} {\bibinfo  {journal} {J. Appl. Crystallogr.}\ }\textbf {\bibinfo
  {volume} {48}},\ \bibinfo {pages} {1560} (\bibinfo {year}
  {2015})}\BibitemShut {NoStop}%
\bibitem [{\citenamefont {Campbell}\ \emph {et~al.}(2006)\citenamefont
  {Campbell}, \citenamefont {Stokes}, \citenamefont {Tanner},\ and\
  \citenamefont {Hatch}}]{Campbell2006}%
  \BibitemOpen
  \bibfield  {author} {\bibinfo {author} {\bibfnamefont {B.~J.}\ \bibnamefont
  {Campbell}}, \bibinfo {author} {\bibfnamefont {H.~T.}\ \bibnamefont
  {Stokes}}, \bibinfo {author} {\bibfnamefont {D.~E.}\ \bibnamefont {Tanner}},
  \ and\ \bibinfo {author} {\bibfnamefont {D.~M.}\ \bibnamefont {Hatch}},\
  }\href {\doibase 10.1107/s0021889806014075607} {\bibfield  {journal}
  {\bibinfo  {journal} {J. Appl. Crystallogr.}\ }\textbf {\bibinfo {volume}
  {39}},\ \bibinfo {pages} {607} (\bibinfo {year} {2006})}\BibitemShut
  {NoStop}%
\bibitem [{\citenamefont {Perez-Mato}\ \emph {et~al.}(2010)\citenamefont
  {Perez-Mato}, \citenamefont {Orobengoa},\ and\ \citenamefont
  {Aroyo}}]{PerezMato2010}%
  \BibitemOpen
  \bibfield  {author} {\bibinfo {author} {\bibfnamefont {J.~M.}\ \bibnamefont
  {Perez-Mato}}, \bibinfo {author} {\bibfnamefont {D.}~\bibnamefont
  {Orobengoa}}, \ and\ \bibinfo {author} {\bibfnamefont {M.~I.}\ \bibnamefont
  {Aroyo}},\ }\href {\doibase 10.1107/S0108767310016247} {\bibfield  {journal}
  {\bibinfo  {journal} {Acta Crystallogr. Sect. A Found. Crystallogr.}\
  }\textbf {\bibinfo {volume} {66}},\ \bibinfo {pages} {558} (\bibinfo {year}
  {2010})}\BibitemShut {NoStop}%
\bibitem [{\citenamefont {Campbell}\ \emph {et~al.}(2007)\citenamefont
  {Campbell}, \citenamefont {Evans}, \citenamefont {Perselli},\ and\
  \citenamefont {Stokes}}]{Campbell2007}%
  \BibitemOpen
  \bibfield  {author} {\bibinfo {author} {\bibfnamefont {B.~J.}\ \bibnamefont
  {Campbell}}, \bibinfo {author} {\bibfnamefont {J.~S.~O.}\ \bibnamefont
  {Evans}}, \bibinfo {author} {\bibfnamefont {F.}~\bibnamefont {Perselli}}, \
  and\ \bibinfo {author} {\bibfnamefont {H.~T.}\ \bibnamefont {Stokes}},\
  }\href {} {\bibfield  {journal} {\bibinfo  {journal} {IUCr Comput. Comm.
  Newsl.}\ }\textbf {\bibinfo {volume} {8}},\ \bibinfo {pages} {81} (\bibinfo
  {year} {2007})}\BibitemShut {NoStop}%
\bibitem [{\citenamefont {Kerman}\ \emph {et~al.}(2012)\citenamefont {Kerman},
  \citenamefont {Campbell}, \citenamefont {Satyavarapu}, \citenamefont
  {Stokes}, \citenamefont {Perselli},\ and\ \citenamefont
  {Evans}}]{Kerman2012}%
  \BibitemOpen
  \bibfield  {author} {\bibinfo {author} {\bibfnamefont {S.}~\bibnamefont
  {Kerman}}, \bibinfo {author} {\bibfnamefont {B.~J.}\ \bibnamefont
  {Campbell}}, \bibinfo {author} {\bibfnamefont {K.~K.}\ \bibnamefont
  {Satyavarapu}}, \bibinfo {author} {\bibfnamefont {H.~T.}\ \bibnamefont
  {Stokes}}, \bibinfo {author} {\bibfnamefont {F.}~\bibnamefont {Perselli}}, \
  and\ \bibinfo {author} {\bibfnamefont {J.~S.~O.}\ \bibnamefont {Evans}},\
  }\href {\doibase 10.1107/S0108767311046241} {\bibfield  {journal} {\bibinfo
  {journal} {Acta Crystallogr. A.}\ }\textbf {\bibinfo {volume} {68}},\
  \bibinfo {pages} {222} (\bibinfo {year} {2012})}\BibitemShut {NoStop}%
\bibitem [{\citenamefont {Levin}\ \emph {et~al.}(2014)\citenamefont {Levin},
  \citenamefont {Krayzman},\ and\ \citenamefont {Woicik}}]{Levin2014}%
  \BibitemOpen
  \bibfield  {author} {\bibinfo {author} {\bibfnamefont {I.}~\bibnamefont
  {Levin}}, \bibinfo {author} {\bibfnamefont {V.}~\bibnamefont {Krayzman}}, \
  and\ \bibinfo {author} {\bibfnamefont {J.~C.}\ \bibnamefont {Woicik}},\
  }\href {\doibase 10.1103/PhysRevB.89.024106} {\bibfield  {journal} {\bibinfo
  {journal} {Phys. Rev. B}\ }\textbf {\bibinfo {volume} {89}},\ \bibinfo
  {pages} {024106} (\bibinfo {year} {2014})}\BibitemShut {NoStop}%
\bibitem [{\citenamefont {Ghosez}\ \emph {et~al.}(1997)\citenamefont {Ghosez},
  \citenamefont {Gonze},\ and\ \citenamefont {Michenaud}}]{Ghosez1997b}%
  \BibitemOpen
  \bibfield  {author} {\bibinfo {author} {\bibfnamefont {P.}~\bibnamefont
  {Ghosez}}, \bibinfo {author} {\bibfnamefont {X.}~\bibnamefont {Gonze}}, \
  and\ \bibinfo {author} {\bibfnamefont {J.~P.}\ \bibnamefont {Michenaud}},\
  }\href {\doibase 10.1080/00150199808009159} {\bibfield  {journal} {\bibinfo
  {journal} {Ferroelectrics}\ }\textbf {\bibinfo {volume} {206}},\ \bibinfo
  {pages} {12} (\bibinfo {year} {1997})}\BibitemShut {NoStop}%
\bibitem [{\citenamefont {Cohen}(1992)}]{Cohen1992}%
  \BibitemOpen
  \bibfield  {author} {\bibinfo {author} {\bibfnamefont {R.~E.}\ \bibnamefont
  {Cohen}},\ }\href {\doibase 10.1038/358136a0} {\bibfield  {journal} {\bibinfo
   {journal} {Nature}\ }\textbf {\bibinfo {volume} {358}},\ \bibinfo {pages}
  {136} (\bibinfo {year} {1992})}\BibitemShut {NoStop}%
\bibitem [{\citenamefont {Coelho}\ \emph {et~al.}(2015)\citenamefont {Coelho},
  \citenamefont {Chater},\ and\ \citenamefont {Kern}}]{Coelho2015}%
  \BibitemOpen
  \bibfield  {author} {\bibinfo {author} {\bibfnamefont {A.~A.}\ \bibnamefont
  {Coelho}}, \bibinfo {author} {\bibfnamefont {P.~A.}\ \bibnamefont {Chater}},
  \ and\ \bibinfo {author} {\bibfnamefont {A.}~\bibnamefont {Kern}},\ }\href
  {\doibase 10.1107/S1600576715007487} {\bibfield  {journal} {\bibinfo
  {journal} {J. Appl. Crystallogr.}\ }\textbf {\bibinfo {volume} {48}},\
  \bibinfo {pages} {869} (\bibinfo {year} {2015})}\BibitemShut {NoStop}%
\end{thebibliography}
\end{document}